\documentclass[12pt]{article}
\begin{document}

\setcounter{page}{1}
\def\theequation{\arabic{section}.\arabic{equation}}
\def\theequation{\thesection.\arabic{equation}}
\setcounter{section}{0}

\title{The relativistic field theory model of the deuteron
from low--energy QCD}

\author{A.N. Ivanov\thanks{E--mail: ivanov@kph.tuwien.ac.at, Tel.:
+43--1--58801--14261, Fax: +43--1--58801--14299}~\thanks{Present
address: The Abdus Salam International Centre for Theoretical Physics ~
(ICTP), {\it Strada Costiera 11, P. O. Box 586, 34100 Trieste,
Italy}}~$^\S$, H. Oberhummer\thanks{E--mail: ohu@kph.tuwien.ac.at,
Tel.: +43--1--58801--14251, Fax: +43--1--58801--14299} ,
N.I. Troitskaya$^\dagger$\thanks{Permanent address: State Technical
University, Department of Nuclear Physics, 195251 St.Petersburg} ,\\
M. Faber\thanks{E--mail: faber@kph.tuwien.ac.at, Tel.:
+43--1--58801--14261, Fax: +43--1--58801--14299}}

\date{\today}
\maketitle
\vspace{-0.5in}

\begin{center}
{\it Institut f\"ur Kernphysik, Technische Universit\"at Wien,\\
Wiedner Hauptstr. 8--10/142, A--1040, Vienna, Austria}
\end{center}

\begin{abstract}
The relativistic field theory model of the deuteron (RFMD) is
reformulated from the first principles of QCD. The deuteron appears as
a neutron--proton collective excitation, i.e. a Cooper np--pair,
induced by a phenomenological local four--nucleon interaction in the
nuclear phase of QCD. The RFMD describes the deuteron coupled to
hadrons through one--nucleon loop exchanges providing a minimal
transfer of nucleon flavours from initial to final nuclear states and
accounting for contributions of nucleon--loop anomalies which are
completely determined by one--nucleon loop diagrams. The dominance of
contributions of nucleon--loop anomalies to effective Lagrangians of
low--energy nuclear interactions is justified in the large $N_C$
expansion, where $N_C$ is the number of quark colours.
\end{abstract}

\begin{center}
\end{center}
\newpage

\section{Introduction}

\hspace{0.2in} In our recent publications [1--5] we have formulated
the relativistic field theory model of the deuteron (RFMD) [1,2]. The
first version of the RFMD [1,2] formulated in analogy with a
phenomenological $\sigma$--model had no obvious connection with QCD
[2]. Unlike the potential model approach (PMA) [6--8] and the
Effective Field Theory (EFT) approach [9--14] the RFMD takes into
account non--perturbative contributions of high--energy (or
short--distance) fluctuations of virtual nucleon $(N)$ and
anti--nucleon $(\bar{N})$ fields, i.e. $N\bar{N}$ fluctuations, in the
form of anomalies of one--nucleon loop diagrams\footnote{In Ref.[5] we
have considered a modified version of the RFMD which is not well
defined due to a violation of Lorentz invariance of the effective
four--nucleon interaction describing N + N $\to$ N + N
transitions. This violation has turned out to be incompatible with a
dominance of one--nucleon loop anomalies which are Lorentz
covariant. Thereby, the astrophysical factor for the solar proton
burning calculated in Ref.[5] and  enhanced by a factor of 1.4 with respect
to the recommended value [7] is not good established.}. The
description of one--nucleon loop anomalies caused by contributions of
high--energy (or short--distance) fluctuations of virtual nucleon
fields goes beyond the scope of both the PMA and the EFT
approach. This is due to the absence in these approaches anti--nucleon
degrees of freedom related to the non--perturbative quantum vacuum --
the nucleon Dirac sea [15].
 
It is known that the nucleon Dirac sea cannot be ignored fully in the
low--energy nuclear physics.  For example, high--energy $N\bar{N}$
fluctuations of the nucleon Dirac sea polarized by the nuclear medium
decrease the scalar nuclear density in the nuclear interior of finite
nuclei by 15$\%$ [16]. This effect has been obtained within quantum
field theoretic approaches in terms of one-nucleon loop
exchanges. Unfortunately, contributions of nucleon--loop anomalies
have not been taken into account in these approaches. The RFMD allows
to fill this blank. In fact, in accord the analysis carried out in
Refs.\,[17] nucleon--loop anomalies can be interpreted as non--trivial
contributions of the nucleon Dirac sea.

In this paper we change the starting ideas of the RFMD having been
formulated in [1,2]. We show that the RFMD is fully motivated by QCD
and describes low--energy nuclear interactions in the nuclear phase of
QCD through one--nucleon loop exchanges. Within the large $N_C$
expansion [18,19] we justify the dominance of one--nucleon loop anomaly
contributions to effective Lagrangians describing the deuteron itself
and processes of low--energy interactions of the deuteron coupled to
other particles.

The paper is organized as follows. In Sect.\,2 we discuss
non--perturbative phases of QCD and formulate the RFMD from the first
principles of QCD. In Sect.\,3 we derive the effective Lagrangian for
the deuteron field induced in the nuclear phase of QCD as the
neutron--proton collective excitation (the Cooper np--pair) by a
phenomenological local four--nucleon interaction.  We demonstrate the
dominant role of one--nucleon loop anomalies for the formation of the
effective Lagrangian of the physical deuteron field. In Sect.\,4 we
investigate the electromagnetic properties of the deuteron and derive
the effective Lagrangian of the deuteron field, the Cooper np--pair,
coupled to an external electromagnetic field describing the magnetic
dipole and electric quadrupole moments of the physical deuteron. In
the Conclusion we discuss the obtained results.

\section{Non--perturbative phases of QCD}
\setcounter{equation}{0}

\hspace{0.2in} The derivation of the RFMD from the first principles of
QCD goes through three non--perturbative phases of the
quark--gluon system. We call them as: 1) the low--energy quark--gluon
phase (low--energy QCD), 2) the hadronic phase and 3) the nuclear
phase.

{\it The low--energy quark--gluon phase of QCD} can be obtained by
integrating over fluctuations of quark and gluon fields at energies
above the scale of spontaneous breaking of chiral symmetry (SB$\chi$S)
$\Lambda_{\chi} \simeq 1\,{\rm GeV}$.  This results in an effective
field theory, low--energy QCD, describing strong low--energy
interactions of quarks and gluons. The low--energy quark--gluon phase
of QCD characterizes itself by the appearance of low--energy
gluon--field configurations leading to electric colour fluxes
responsible for formation of a linearly rising interquark
potential. The former realizes quark confinement.

For the transition to {\it the hadronic phase of QCD} one should,
first, integrate out low--energy gluon degrees of freedom.
Integrating over gluon degrees of freedom fluctuating around
low--energy gluon--field configurations responsible for formation of a
linearly rising interquark potential one arrives at an effective field
theory containing only quark ($q$) and anti--quark ($\bar{q}$) degrees
of freedom. This effective field theory describes strong interactions
at energies below the SB$\chi$S scale $\Lambda_{\chi} \simeq 1\,{\rm
GeV}$.  The resultant quark system possesses both a chirally invariant
and a chirally broken phase. In the chirally invariant phase the
effective Lagrangian of the quark system is invariant under chiral
$U(3)\times U(3)$ group. The chirally invariant phase of the quark
system is unstable and the transition to the chirally broken phase is
advantageous. The chirally broken phase characterizes itself by three
non--perturbative phenomena: SB$\chi$S, hadronization (creation of
bound quark states with quantum numbers of mesons $q\bar{q}$,
$qq\bar{q}\bar{q}$, baryons $qqq$ and so on) and confinement. The
transition to the chirally broken phase caused by SB$\chi$S
accompanies itself with hadronization. Due to quark confinement all
observed bound quark states should be colourless. As in such an
effective field theory gluon degrees of freedom are integrated out,
the entire variety of strong low--energy interactions of hadrons at
energies below the SB$\chi$S scale $\Lambda_{\chi}\simeq 1\,{\rm GeV}$
is described by quark--loop exchanges.

Since nowadays in continuum space--time formulation of QCD the
integration over low--energy gluon--field configurations can be hardly
performed explicitly, phenomenological approximations of this
integration represented by different effective quark models with
chiral $U(3)\times U(3)$ symmetry motivated by QCD are welcomed.

The most interesting effective quark model allowing to describe
analytically both SB$\chi$S and bosonization (creation of bound
$q\bar{q}$ states with quantum numbers of observed low--lying mesons)
is the extended Nambu--Jona--Lasinio (ENJL) model [20--24] with linear
[20,22] and non--linear [21,23] realization of chiral $U(3)\times
U(3)$ symmetry. As has been shown in Ref.\,[24] the ENJL model is fully
motivated by low--energy QCD with a linearly rising interquark
potential and $N_C$ quark colour degrees of freedom at $N_C \to
\infty$. In the ENJL model mesons are described as $q\bar{q}$
collective excitations (the Cooper $q\bar{q}$--pairs) induced by
phenomenological local four--quark interactions. Through
one--constituent quark--loop exchanges the Copper $q\bar{q}$--pairs
acquire the properties of the observed low--lying mesons such as
$\pi(140)$, $K(490)$, $\eta(550)$, $\rho(770)$, $\omega(780)$,
$K^*(890)$ and so on. For the description of low--lying octet and
decuplet of baryons the ENJL model has been extended by the inclusion
of local six--quark interactions responsible for creation of baryons
as $qqq$ collective excitations [25].

Integrating then out low--energy quark--field fluctuations, that can
be performed in terms of constituent quark--loop exchanges, one
arrives at {\it the hadronic phase of QCD} containing only local meson
and baryon fields. The couplings of low--lying mesons and baryons are
described by Effective Chiral Lagrangians with chiral $U(3)\times
U(3)$ symmetry [20--27].

{\it The nuclear phase of QCD} characterizes itself by the appearance
of bound nucleon states -- nuclei. In order to arrive at {\it the
nuclear phase of QCD} we suggest to start with the hadronic phase of
QCD and integrate out heavy hadron degrees of freedom, i.e. all heavy
baryon degrees of freedom with masses heavier than masses of
low--lying octet and decuplet of baryons and heavy meson degrees of
freedom with masses heavier than the SB$\chi$S scale $\Lambda_{\chi}
\simeq 1\,{\rm GeV}$.  At low energies the result of the integration
over these heavy hadron degrees of freedom can be represented in the
form of phenomenological local many--nucleon interactions. Following
the scenario of the hadronic phase of QCD, where hadrons are produced
by phenomenological local many--quark interactions as many--quark
collective excitations, one can assume that some of these
many--nucleon interactions are responsible for creation of
many--nucleon collective excitations. These excitations acquire the
properties of observed bound nucleon states -- nuclei through
nucleon--loop and low--lying meson exchanges. This results in an
effective field theory describing nuclei and their low--energy
interactions in analogy with Effective Chiral Lagrangian approaches
[26,27].  Note that Chiral perturbation theory  can be  naturally
incorporated into this effective field theory of low--energy
interactions of nuclei.

We would like to emphasize that in this scenario of the quantum field
theoretic formation of nuclei and their low--energy interactions
nuclei are considered as elementary particles. For the first time, the
representation of nuclei as elementary particles has been suggested by
Sakita and Goebal [28] and Kim and Primakoff [29] for the description
of electromagnetic and weak nuclear processes. We develop the quantum
field theoretic approach to the interpretation of nuclei as elementary
particles by starting with QCD.

Following this scenario the deuteron, being the lightest bound nucleon
state, appears in the nuclear phase of QCD as the neutron--proton
collective excitation (the Cooper np--pair) induced by a
phenomenological local four--nucleon interaction. Through one--nucleon
loop exchanges [1--5,20--24] the Cooper np--pair with quantum
numbers of the physical deuteron acquires the properties of the
physical deuteron (i) the binding energy $\varepsilon_{\rm D} =
2.225\,{\rm MeV}$, (ii) the electric quadrupole moment $Q_{\rm D} =
0.286\,{\rm fm}^2$ [30] and so on.

\section{The deuteron as a Cooper np--pair}
\setcounter{equation}{0}

\hspace{0.2in} In order to describe the deuteron induced as the Cooper
np--pair we introduce a phenomenological local four--nucleon
interaction caused by the integration over heavy hadron degrees of
freedom
\begin{eqnarray}\label{label3.1}
{\cal L}_{\rm int}(x) = - \frac{g^2_{\rm V}}{4M^2_{\rm
N}}\,j^{\dagger}_{\mu}(x)j^{\mu}(x),
\end{eqnarray}
where $g_{\rm V}$ is the phenomenological coupling constant of the
RFMD [1--4], $M_{\rm N} = 940\,{\rm MeV}$ is the nucleon mass. We
neglect here the electromagnetic mass difference for the neutron and
the proton.  As has been found in [1,2] the coupling constant $g_{\rm
V}$ is related to the electric quadrupole moment of the deuteron
$Q_{\rm D}$: $g^2_{\rm V}=2 \pi^2 Q_{\rm D}M^2_{\rm N}$ [2].

The baryon current $j^{\mu}(x)$ with the quantum numbers of
the deuteron is defined by [1--4]
\begin{eqnarray}\label{label3.2}
j^{\mu}(x)= -i\,[\bar{p^c}(x)\gamma^{\mu}n(x) -
\bar{n^c}(x)\gamma^{\mu}p(x)].
\end{eqnarray}
Here $p(x)$ and $n(x)$ are the interpolating fields of the proton and
the neutron, $N^c(x) = C\,\bar{N}^T(x)$ and $\bar{N^c}(x) =
N^T(x)\,C$, where $C$ is a charge conjugation matrix and $T$ is a
transposition.  In terms of the electric quadrupole moment of the
deuteron the phenomenological local four--nucleon interaction
Eq.(\ref{label3.1}) reads
\begin{eqnarray}\label{label3.3}
{\cal L}_{\rm int}(x) = - \frac{1}{2}\,\pi^2\,Q_{\rm
D}\,j^{\dagger}_{\mu}(x)j^{\mu}(x).
\end{eqnarray}
Now let us discuss the behaviour of the phenomenological coupling
constant $g^2_{\rm V}/4M^2_{\rm N}$ from the point of view of the
large $N_C$ expansion in QCD with the $SU(N_C)$ gauge group at $N_C
\to \infty$ [18,19]. Suppose, for simplicity, that the
phenomenological four--nucleon interaction Eq.(\ref{label3.1}) is
caused by exchanges of the scalar $f_0(980)$ and $a_0(980)$ mesons
being the lightest states among heavy hadrons we have integrated out.

Through a linear realization of chiral $U(3)\times U(3)$ symmetry and
the Goldberger--Treiman relation one can find that the coupling
constant of $\sigma$--mesons $g_{\rm \sigma NN}$, the
$q\bar{q}$--scalar mesons, coupled to the octet of low--lying baryons
should be of order $g_{\rm \sigma NN} \sim O(\sqrt{N_C})$ at $N_C \to
\infty$. The scalar mesons $f_0(980)$ and $a_0(980)$ are most likely
four--quark states with $qq\bar{q}\bar{q}$ quark structure [31,32]. In
the limit $N_C \to \infty$ such $qq\bar{q}\bar{q}$ states are
suppressed by a factor $1/N_C$ [19]. Thus, an effective coupling
constant of low--energy NN interaction caused by the
$qq\bar{q}\bar{q}$ scalar meson exchanges should be of order
$O(1/N_C)$ at $N_C \to \infty$. By taking into account that in QCD
with $N_C \to \infty$ the nucleon mass $M_{\rm N}$ is proportional to
$N_C$ [19], $M_{\rm N} = N_C M_q $, where $M_q \sim 300\,{\rm MeV}$ is
the constituent quark mass, we can introduce the nucleon mass $M_{\rm
N}$ in the effective coupling constant as a dimensional parameter
absorbing the factor $N^2_C$, i.e. $g^2_{\rm V}/4M^2_{\rm N}$.  This
is also required by the correct dependence of the deuteron mass on
$N_C$.  As a result the phenomenological coupling constant $g_{\rm V}$
turns out to be of order $O(\sqrt{N_C})$ at $N_C \to \infty$.

We should emphasize that one does not need to know too much about
quark structure of heavy hadron degrees of freedom we have integrated
out. Without loss of generality one can argue that among the multitude
of contributions caused by the integration over heavy hadron degrees
of freedom one can always find the required local four--nucleon
interaction the effective coupling constant of which behaves like
$O(1/N_C)$ at $N_C \to \infty$. As we show below this behaviour of the
coupling constant of the phenomenological four--nucleon interaction is
consistent with the large $N_C$ dependence of low--energy parameters
of the physical deuteron.

The effective Lagrangian of the np--system unstable under creation of
the Cooper np--pair with quantum numbers of the deuteron
is then defined
\begin{eqnarray}\label{label3.4}
{\cal L}^{\rm np}(x)&=&\bar{n}(x)\,(i\gamma^{\mu}\partial_{\mu} - M_{\rm
N})\,n(x)+ \bar{p}(x)\,(i\gamma^{\mu}\partial_{\mu} -
M_{\rm N})\,p(x)  \nonumber\\ 
&&- \frac{g^2_{\rm V}}{4M^2_{\rm
N}}\,j^{\dagger}_{\mu}(x)j^{\mu}(x) + \ldots,
\end{eqnarray}
where ellipses stand for low--energy interactions the neutron and the
proton with other fields.  

In order to introduce the interpolating local deuteron field we should
linearalize the Lagrangian Eq.(\ref{label3.4}). Following the
procedure described in Refs.\,[20--23] for the inclusion of local
interpolating meson fields in the ENJL model we get
\begin{eqnarray}\label{label3.5}
&&{\cal L}^{\rm np}(x) \to \bar{n}(x)\,(i\gamma^{\mu}\partial_{\mu} -
M_{\rm N})\,n(x) + \bar{p}(x)\,(i\gamma^{\mu}\partial_{\mu} - M_{\rm
N})\,p(x) \nonumber\\ 
&&+M^2_0\,D^{\dagger}_{\mu}(x)D^{\mu}(x) +
g_{\rm V}j^{\dagger}_{\mu}(x)D^{\mu}(x) + g_{\rm
V}j^{\mu}(x)D^{\dagger}_{\mu}(x) + \ldots,
\end{eqnarray}
where $M_0 = 2\,M_{\rm N}$ and $D^{\mu}(x)$ is a local
interpolating field with quantum numbers of the deuteron.

In order to derive the effective Lagrangian of the physical deuteron
field we should integrate over nucleon fields in the one--nucleon loop
approximation [1--5,20--24]. The one--nucleon loop approximation of
low--energy nuclear forces allows (i) to transfer nucleon flavours
from an initial to a final nuclear state by a minimal way and (ii) to
take into account contributions of nucleon--loop anomalies [1--5,17,
33--35], which are fully defined by one--nucleon loop diagrams
[33--35]. It is well--known that quark--loop anomalies play an
important role for the correct description of strong low--energy
interactions of low--lying hadrons [20--27]. We argue the dominant
role of nucleon--loop anomalies for the correct description of
low--energy nuclear forces in nuclear physics. We demonstrate below
the dominant role of nucleon--loop anomalies by example of the
evaluation of the effective Lagrangian of the free deuteron field.

The effective Lagrangian of the free deuteron field evaluated in the
one--nucleon loop approximation is defined by [1--4]:
\begin{eqnarray}\label{label3.6}
\hspace{-0.5in}&&\int d^4x\,{\cal L}_{\rm eff}(x) =\int
d^4x\,M^2_0\,D^{\dagger}_{\mu}(x)D^{\mu}(x) \nonumber\\
\hspace{-0.5in}&& - \int d^4x\int\frac{d^4x_1
d^4k_1}{(2\pi)^4}\,e^{\textstyle -ik_1\cdot(x -
x_1)}\,D^{\dagger}_{\mu}(x) D_{\nu}(x_1)\,\frac{g^2_{\rm
V}}{4\pi^2}\,\Pi^{\mu\nu}(k_1; Q),
\end{eqnarray}
where the structure function $\Pi^{\mu\nu}(k_1; Q)$ is given by
\begin{eqnarray}\label{label3.7}
\Pi^{\mu\nu}(k_1; Q)=\int\frac{d^4k}{\pi^2i}{\rm
tr}\Bigg\{\frac{1}{M_{\rm N} - \hat{k} - \hat{Q} -
\hat{k}_1}\gamma^{\mu}\frac{1}{M_{\rm N} - \hat{k} -
\hat{Q}}\gamma^{\nu}\Bigg\}.
\end{eqnarray}
The 4--momentum $Q = a\,k_1$ is an arbitrary shift of momenta of
virtual nucleon fields with an arbitrary parameter $a$ [1,2]. For
the evaluation of the $Q$--dependence of the structure function
$\Pi^{\mu\nu}(k_1; Q)$ we apply the procedure suggested by Gertsein
and Jackiw [32] (see also [2]):
\begin{eqnarray}\label{label3.8}
&&\Pi^{\mu\nu}(k_1; Q) - \Pi^{\mu\nu}(k_1; 0)
=\int\limits^1_0dx\,\frac{d}{dx}\,\Pi^{\mu\nu}(k_1; x Q)=\nonumber\\
&&=\int\limits^1_0dx\int
\frac{d^4k}{\pi^2i}Q^{\lambda}\frac{\partial}{\partial
k^{\lambda}}{\rm tr}\Bigg\{\frac{1}{M_{\rm N} - \hat{k} - x\hat{Q} -
\hat{k}_1}\gamma^{\mu}\frac{1}{M_{\rm N} - \hat{k} -
x\hat{Q}}\gamma^{\nu}\Bigg\}=\nonumber\\ 
&&= 2 \int\limits^1_0
dx\,\lim_{k\to \infty}\Bigg<\frac{Q\cdot k}{k^2}{\rm tr}\{(M_{\rm N} +
\hat{k} + x\hat{Q} + \hat{k}_1)\gamma^{\mu}(M_{\rm N} + \hat{k} +
x\hat{Q})\gamma^{\nu}\}\Bigg>=\nonumber\\ 
&&=2\,(2Q^{\mu}Q^{\nu} -
Q^2\,g^{\mu\nu}) + 2(k^{\mu}_1Q^{\nu} + k^{\nu}_1Q^{\mu} - k_1\cdot
Q\,g^{\mu\nu})=\nonumber\\ &&= -2\,a(a+1)\,(k^2_1\,g^{\mu\nu} -
2\,k^{\mu}_1k^{\nu}_1).
\end{eqnarray}
Thus, we obtain
\begin{eqnarray}\label{label3.9}
\Pi^{\mu\nu}(k_1; Q) -
\Pi^{\mu\nu}(k_1; 0)= -2\,a(a+1)\,(k^2_1\,g^{\mu\nu} -
2\,k^{\mu}_1k^{\nu}_1).
\end{eqnarray}
We would like to emphasize that the r.h.s. of Eq.(\ref{label3.9}) is
an explicit expression completely defined by high--energy
(short--distance) $N\bar{N}$ fluctuations, since the virtual momentum
$k$ is taken at the limit $k \to \infty$, and related to the anomaly
of the one--nucleon loop diagram with two vector vertices (the
VV--diagram) [33,35].

The structure function $\Pi^{\mu\nu}(k_1; 0)$ has been evaluated in
Refs.~[1,2,5] and reads
\begin{eqnarray}\label{label3.10}
\Pi^{\mu\nu}(k_1; 0)=\frac{4}{3}(k^2_1g^{\mu\nu} -
k^{\mu}_1k^{\nu}_1)J_2(M_{\rm
N}) + 2g^{\mu\nu}[J_1(M_{\rm N}) + M^2_{\rm N}J_2(M_{\rm
N})],
\end{eqnarray}
where  $J_1(M_{\rm N})$ and $J_2(M_{\rm N})$ are
the quadratically and logarithmically divergent integrals [1,2,5]:
\begin{eqnarray}\label{label3.11}
J_1(M_{\rm N})&=&\int\frac{d^4k}{\pi^2i}\frac{1}{M^2_{\rm N} - k^2} =
4 \int\limits^{\textstyle \Lambda_{\rm
D}}_0\frac{d|\vec{k}\,|\vec{k}^{\,2}}{(M^2_{\rm N}
+\vec{k}^{\,2})^{1/2}},\nonumber\\ 
J_2(M_{\rm
N})&=&\int\frac{d^4k}{\pi^2i}\frac{1}{(M^2_{\rm N} - k^2)^2} = 2
\int\limits^{\textstyle \Lambda_{\rm
D}}_0\frac{d|\vec{k}\,|\vec{k}^{\,2}}{(M^2_{\rm N}
+\vec{k}^{\,2})^{3/2}}.
\end{eqnarray}
The cut--off $\Lambda_{\rm D}$ restricts from above 3--momenta of
low--energy fluctuations of virtual neutron and proton fields forming
the physical deuteron [1--5]. Since in the RFMD the cut--off
$\Lambda_{\rm D}$ is much less than the mass of the nucleon, $M_{\rm
N} \gg \Lambda_{\rm D}$ [1,2], we use below the relation between the
divergent integrals:
\begin{eqnarray}\label{label3.12}
J_1(M_{\rm N})= 2\,M^2_{\rm N}\,J_2(M_{\rm N}) =
\frac{4}{3}\,\frac{\Lambda^3_{\rm D}}{M_{\rm N}} \sim O(1/N_C).
\end{eqnarray}
Note that in Eq.(\ref{label3.10}) we have taken into account only the
leading terms in the external momentum expansion, i.e. the
$k_1$--expansion.

The justification of the dominance of the leading order contributions
in expansion in powers of external momenta can be provided in the
large $N_C$ expansion. Indeed, in QCD with the $SU(N_C)$ gauge group
at $N_C \to \infty$ the baryon mass is proportional to the number of
quark colours [19]: $M_{\rm N} \sim N_C$.  Since for the derivation of
effective Lagrangians describing the deuteron and amplitudes of
processes of low--energy interactions of the deuteron coupled to other
particles all external momenta of interacting particles should be kept
off--mass shell, the masses of virtual nucleon fields taken at $N_C
\to \infty$ are larger compared with external momenta. By expanding
one--nucleon loop diagrams in powers of $1/M_{\rm N}$ we get an
expansion in powers of $1/N_C$. Keeping the leading order in the large
$N_C$ expansion we are leaving with the leading order contributions in
an external momentum expansion. We should emphasize that anomalous
contributions of one--nucleon loop diagrams are defined by the least
powers of an external momentum expansion. This implies that in the
RFMD effective Lagrangians of low--energy interactions are completely
determined by contributions of one--nucleon loop anomalies. The
divergent contributions having the same order in momentum expansion
are negligible small compared with the anomalous ones due to the
inequality $M_{\rm N} \gg \Lambda_{\rm D}$ [1--5] and the limit $N_C
\to \infty$. This justifies the application of the approximation by
the leading powers in an external momentum expansion to the evaluation
of the effective Lagrangians of the deuteron coupled to itself and an
external electromagnetic field [2], and the effective Lagrangians
describing amplitudes of low--energy nuclear processes like the solar
proton burning p + p $\to$ D + e$^+$ + $\nu_{\rm e}$ and so [2,4].

Collecting all pieces we get the structure function $\Pi^{\mu\nu}(k_1;
Q)$ in the form
\begin{eqnarray}\label{label3.13}
\Pi^{\mu\nu}(k_1; Q)&=&\frac{4}{3}(k^2_1g^{\mu\nu} -
k^{\mu}_1k^{\nu}_1)J_2(M_{\rm
N}) + 2g^{\mu\nu}[J_1(M_{\rm N}) + M^2_{\rm N}J_2(M_{\rm
N})]\nonumber\\
&& -  2\,a(a+1)\,(k^2_1\,g^{\mu\nu} -
2\,k^{\mu}_1k^{\nu}_1).
\end{eqnarray}
The effective  Lagrangian of the free deuteron field is then defined
\begin{eqnarray}\label{label3.14}
&&{\cal L}_{\rm eff}(x)= -\frac{1}{2}\Bigg(-\frac{g^2_{\rm
V}}{2\pi^2}\,a(a+1)+ \frac{g^2_{\rm
V}}{3\pi^2}\,J_2(M_{\rm
N})\Bigg)\,D^{\dagger}_{\mu\nu}(x)D^{\mu\nu}(x)\nonumber\\
&& + \Bigg(M^2_0 - \frac{g^2_{\rm
V}}{2\pi^2}\,[J_1(M_{\rm N}) + M^2_{\rm N}J_2(M_{\rm
N})]\Bigg)\,D^{\dagger}_{\mu}(x)D^{\mu}(x),
\end{eqnarray}
where $D^{\mu\nu}(x)=\partial^{\mu}D^{\nu}(x) -
\partial^{\nu}D^{\mu}(x)$. We have dropped some  contributions
proportional to the total divergence of the deuteron field, since
$\partial_{\mu}D^{\mu}(x) = 0$. For the derivation of
Eq.(\ref{label3.14}) we have used the relation
\begin{eqnarray}\label{label3.15}
&&\int d^4x\int\frac{d^4x_1 d^4k_1}{(2\pi)^4}\,e^{\textstyle
-ik_1\cdot(x - x_1)}\,D^{\dagger}_{\mu}(x)D_{\nu}(x_1)(k^2_1g^{\mu\nu}
- k^{\mu}_1k^{\nu}_1)=\nonumber\\ 
&&= \frac{1}{2}\int
d^4x\,D^{\dagger}_{\mu\nu}(x)D^{\mu\nu}(x).
\end{eqnarray}
In order to get a correct kinetic term of the free deuteron field in
the effective Lagrangian Eq.(\ref{label3.14}) we should set [2]
\begin{eqnarray}\label{label3.16}
-\frac{g^2_{\rm
V}}{2\pi^2}\,a(a+1)=1.
\end{eqnarray}
Since $a$ is an arbitrary real parameter, the relation
Eq.(\ref{label3.16}) is valid in the case of the existence of real
roots.  For the existence of real roots of Eq.(\ref{label3.16}) the
coupling constant $g_{\rm V}$ should obey the constraint $g^2_{\rm V}>
8\pi^2$ that is satisfied by the numerical value $g_{\rm V} = 11.319$
calculated at $N_C = 3$ [2]. Since $g_{\rm V} \sim O(\sqrt{N_C})$ at
$N_C \to \infty$, Eq.(\ref{label3.16}) has real solutions for any $N_C
\ge 3$.

Due to Eq.(\ref{label3.16}) the effective Lagrangian of the free
deuteron field takes the form
\begin{eqnarray}\label{label3.17}
&&{\cal L}_{\rm eff}(x)= -\frac{1}{2}\Bigg(1 + \frac{g^2_{\rm
V}}{3\pi^2}\,J_2(M_{\rm
N})\Bigg)\,D^{\dagger}_{\mu\nu}(x)D^{\mu\nu}(x)\nonumber\\
&& + \Bigg(M^2_0 - \frac{g^2_{\rm
V}}{2\pi^2}\,[J_1(M_{\rm N}) + M^2_{\rm N}J_2(M_{\rm
N})]\Bigg)\,D^{\dagger}_{\mu}(x)D^{\mu}(x).
\end{eqnarray}
By performing the renormalization of the wave function of the deuteron
field [1,2]
\begin{eqnarray}\label{label3.18}
\Bigg(1 + \frac{g^2_{\rm
V}}{3\pi^2}\,J_2(M_{\rm
N})\Bigg)^{1/2}\,D^{\mu}(x) \to D^{\mu}(x)
\end{eqnarray}
and taking into account that $M_{\rm N} \gg \Lambda_{\rm D}$ we arrive
at the effective Lagrangian of the free physical deuteron field
\begin{eqnarray}\label{label3.19}
{\cal L}_{\rm eff}(x)=
-\frac{1}{2}\,D^{\dagger}_{\mu\nu}(x)D^{\mu\nu}(x) + M^2_{\rm
D}\,D^{\dagger}_{\mu}(x)D^{\mu}(x),
\end{eqnarray}
where $M_{\rm D} = M_0 - \varepsilon_{\rm D}$ is the mass of the
physical deuteron field. The binding energy of the deuteron
$\varepsilon_{\rm D}$ reads 
\begin{eqnarray}\label{label3.20}
\varepsilon_{\rm D}= \frac{17}{48}\,\frac{g^2_{\rm
V}}{\pi^2}\,\frac{J_1(M_{\rm N})}{M_{\rm N}}= \frac{17}{18}\,Q_{\rm
D}\,\Lambda^3_{\rm D} \sim O(1/N_C).
\end{eqnarray}
We have used here the relation between divergent integrals
Eq.(\ref{label3.11}) and expressed the phenomenological coupling
constant $g_{\rm V}$ in terms of the electric quadrupole moment of the
deuteron $g^2_{\rm V} = 2\pi^2Q_{\rm D}M^2_{\rm N}$.

At $N_C \to \infty$ the binding energy of the deuteron behaves like
$O(1/N_C)$ as well as the electric quadrupole moment $Q_{\rm D}$ and
the coupling constant of the phenomenological local four--nucleon
interaction Eq.(\ref{label3.1}). This testifies a self--consistency of
our approach. Really, all parameters of the physical deuteron field
are of the same order according to the large $N_C$ expansion. This
means that the vanishing of the coupling constant of the
phenomenological four--nucleon interaction Eq.(\ref{label3.1}) in the
limit $N_C \to \infty$ entails the vanishing of all low--energy
parameters of the physical deuteron.

\section{Electromagnetic properties of the deuteron}
\setcounter{equation}{0}

The description of the deuteron as a Cooper np--pair changes a little
bit the description of the electromagnetic parameters of the deuteron
given in Ref.[2]. We do not have more a ``bare'' deuteron field having
magnetic dipole and electric quadrupole moments. Therefore, both the
magnetic dipole and electric quadrupole moments have to be described
by the one--nucleon loop contributions. For the self--consistent
description of the electromagnetic properties of the deuteron we
cannot deal with only the baryon current $j_{\mu}(x)$ given by
Eq.(\ref{label3.2}) and have to introduce the tensor current [2] 
\begin{eqnarray}\label{label4.1}
{\cal J}^{\mu\nu}(x)= \bar{p^c}(x)\sigma^{\mu\nu}n(x) -
\bar{n^c}(x)\sigma^{\mu\nu}p(x),
\end{eqnarray}
where $\sigma^{\mu\nu}=(\gamma^{\mu}\gamma^{\nu}-
\gamma^{\nu}\gamma^{\mu})/2$.

The local four--nucleon interaction producing the deuteron as a Cooper
np--pair reads now 
\begin{eqnarray}\label{label4.2}
{\cal L}_{\rm int}(x) = - \frac{1}{4M^2_{\rm
N}}\,J^{\dagger}_{\mu}(x)J^{\mu}(x).
\end{eqnarray}
The baryon current $J^{\mu}(x)$ is defined by
\begin{eqnarray}\label{label4.3}
J^{\mu}(x) &=& -i\,g_{\rm V}\,[\bar{p^c}(x)\gamma^{\mu}n(x) -
\bar{n^c}(x)\gamma^{\mu}p(x)]\nonumber\\
&&-\,\frac{g_{\rm T}}{2M_{\rm
N}}\,\partial_{\nu}[\bar{p^c}(x)\sigma^{\nu\mu}n(x) -
\bar{n^c}(x)\sigma^{\nu\mu}p(x)],
\end{eqnarray}
where $g_{\rm T}$ is a dimensionless phenomenological coupling
constant [2]. The contribution of the tensor nucleon current looks
like the next--to--leading term in the long--wavelength
expansion\footnote{Due to proportionality $M_{\rm N} \sim N_C$ this
expansion is related to the large $N_C$ expansion.} of an effective
low--energy four--nucleon interaction.

The effective Lagrangian of the np--system unstable under creation of
the Cooper np--pair with quantum numbers of the deuteron
is then defined
\begin{eqnarray}\label{label4.4}
{\cal L}^{\rm np}(x)&=&\bar{n}(x)\,(i\gamma^{\mu}\partial_{\mu} - M_{\rm
N})\,n(x)+ \bar{p}(x)\,(i\gamma^{\mu}\partial_{\mu} -
M_{\rm N})\,p(x)  \nonumber\\ 
&&- \frac{1}{4M^2_{\rm
N}}\,J^{\dagger}_{\mu}(x)J^{\mu}(x).
\end{eqnarray}
The linearalized version of the effective Lagrangian
Eq.(\ref{label4.4}) containing the interpolating local deuteron field
reads
\begin{eqnarray}\label{label4.5}
\hspace{-0.5in}{\cal L}^{\rm np}(x)&\to& \bar{n}(x)\,(i\gamma^{\mu}\partial_{\mu} - M_{\rm
N})\,n(x) + \bar{p}(x)\,(i\gamma^{\mu}\partial_{\mu} - M_{\rm
N})\,p(x)\nonumber\\
\hspace{-0.5in}&&  + M^2_0\,D^{\dagger}_{\mu}(x)D^{\mu}(x) + g_{\rm V}j^{\dagger}_{\mu}(x)D^{\mu}(x) + g_{\rm
V}j^{\mu}(x)D^{\dagger}_{\mu}(x)\nonumber\\
\hspace{-0.5in}&& + \frac{g_{\rm T}}{M_0}J^{\dagger}_{\mu\nu}(x)
D^{\mu\nu}(x) + \frac{g_{\rm T}}{M_0}J^{\mu\nu}(x)
D^{\dagger}_{\mu\nu}(x),
\end{eqnarray}
where $M_0 = 2\,M_{\rm N}$, $D^{\mu}(x)$ is a local interpolating
field with quantum numbers of the deuteron and $D^{\mu\nu}(x) =
\partial^{\mu} D^{\nu}(x) - \partial^{\nu} D^{\mu}(x)$.

The interactions with the tensor current give the contributions only
to the divergent part of the effective Lagrangian of the free deuteron
field defined now by [2]
defined
\begin{eqnarray}\label{label4.6}
\hspace{-0.3in}&&{\cal L}_{\rm eff}(x)=
-\frac{1}{2}\Bigg(-\frac{g^2_{\rm V}}{2\pi^2}\,a(a+1)+ \frac{g^2_{\rm
V} + 6g_{\rm V}g_{\rm T} + 3g^2_{\rm T}}{3\pi^2}\,J_2(M_{\rm
N})\Bigg)\,D^{\dagger}_{\mu\nu}(x)D^{\mu\nu}(x)\nonumber\\
\hspace{-0.3in}&&\hspace{0.7in} + \Bigg(M^2_0 - \frac{g^2_{\rm
V}}{2\pi^2}\,[J_1(M_{\rm N}) + M^2_{\rm N}J_2(M_{\rm
N})]\Bigg)\,D^{\dagger}_{\mu}(x)D^{\mu}(x).
\end{eqnarray}
Due to the relation Eq.(\ref{label3.16}) the effective Lagrangian of
the free deuteron field Eq.(\ref{label4.6}) takes the form
\begin{eqnarray}\label{label4.7}
\hspace{-0.3in}&&{\cal L}_{\rm eff}(x)=
-\frac{1}{2}\Bigg(1 + \frac{g^2_{\rm
V} +  6g_{\rm V}g_{\rm T} + 3g^2_{\rm T}}{3\pi^2}\,J_2(M_{\rm
N})\Bigg)\,D^{\dagger}_{\mu\nu}(x)D^{\mu\nu}(x)\nonumber\\
\hspace{-0.3in}&&\hspace{0.7in} + \Bigg(M^2_0 - \frac{g^2_{\rm
V}}{2\pi^2}\,[J_1(M_{\rm N}) + M^2_{\rm N}J_2(M_{\rm
N})]\Bigg)\,D^{\dagger}_{\mu}(x)D^{\mu}(x).
\end{eqnarray}
After the renormalization of the wave function of the deuteron field
we arrived at the effective Lagrangian defined by Eq.(\ref{label3.19})
with the binding energy of the deuteron depending on $g_{\rm V}$ and
$g_{\rm T}$ [2]
\begin{eqnarray}\label{label4.8}
\varepsilon_{\rm D} &=& \frac{17}{48}\,\frac{g^2_{\rm
V}}{\pi^2}\,\frac{J_1(M_{\rm N})}{M_{\rm N}}\,\Bigg( 1 +
\frac{48}{17}\,\frac{g_{\rm T}}{g_{\rm V}} +
\frac{24}{17}\,\frac{g^2_{\rm T}}{g^2_{\rm V}}\Bigg) =\nonumber\\
&=&\frac{17}{18}\,Q_{\rm D}\,\Lambda^3_{\rm D}\,\Bigg( 1 +
\frac{48}{17}\,\frac{g_{\rm T}}{g_{\rm V}} +
\frac{24}{17}\,\frac{g^2_{\rm T}}{g^2_{\rm V}}\Bigg),
\end{eqnarray}
where we have used the relation between divergent integrals
Eq.(\ref{label3.11}) and expressed the phenomenological coupling
constant $g_{\rm V}$ in terms of the electric quadrupole moment of the
deuteron $g^2_{\rm V} = 2\pi^2Q_{\rm D}M^2_{\rm N}$. In order to make
the prediction for the binding energy much more definite we have to
know the relation between the phenomenological coupling constants
$g_{\rm V}$ and $g_{\rm T}$. For this aim we suggest to consider the
electromagnetic properties of the deuteron.

Including the electromagnetic field by a minimal way $\partial_{\mu}
\to \partial_{\mu} + i\,e\,A_{\mu}(x)$, where $e$ and $A_{\mu}(x)$ are
the electric charge of the proton and the electromagnetic potential we
bring up the linearalized version of the Lagrangian
Eq.(\ref{label4.5}) to the form 
\begin{eqnarray}\label{label4.9}
\hspace{-0.5in}&&{\cal L}^{\rm np}(x) \to {\cal L}^{\rm np}_{\rm
ELM}(x) =\nonumber\\
\hspace{-0.5in}&&= \bar{n}(x)\,(i\gamma^{\mu}\partial_{\mu} - M_{\rm
N})\,n(x) + \bar{p}(x)\,(i\gamma^{\mu}\partial_{\mu} - M_{\rm
N})\,p(x) + M^2_0\,D^{\dagger}_{\mu}(x)D^{\mu}(x)\nonumber\\
\hspace{-0.5in}&& + g_{\rm V}j^{\dagger}_{\mu}(x)D^{\mu}(x) + g_{\rm
V}j^{\mu}(x)D^{\dagger}_{\mu}(x) + \frac{g_{\rm
T}}{M_0}J^{\dagger}_{\mu\nu}D^{\mu\nu}(x) + \frac{g_{\rm
T}}{M_0}J^{\mu\nu}D^{\dagger}_{\mu\nu}(x)\nonumber\\ 
\hspace{-0.5in}&&-\,e\,\bar{p}(x)\gamma^{\mu}p(x)\,A_{\mu}(x) - i\,e\,\frac{g_{\rm T}}{M_0}J^{\dagger}_{\mu\nu}(x)(A^{\mu}(x)D^{\nu}(x) -
A^{\nu}(x)D^{\mu}(x))\nonumber\\
\hspace{-0.5in}&& + i\,e\,\frac{g_{\rm
T}}{M_0}J^{\mu\nu}(x)(A_{\mu}(x)D_{\nu}(x) - A^{\nu}(x)D^{\mu}(x)).
\end{eqnarray}
The effective Lagrangian describing both the magnetic dipole and 
electric quadrupole moments of the deuteron has been evaluated in
Ref.[2] and reads
\begin{eqnarray}\label{label4.10}
\delta {\cal L}^{\rm D}_{\rm
ELM}(x)_{\rm eff}&=&i\,e\,\frac{4ag^{2}_{\rm T}-bg^{2}_{\rm
V}}{6\,\pi^{2}}\,D^{\dagger}_{\mu\nu}(x)\,A^{\mu}(x)\,D^{\nu}(x)\nonumber\\
&-&i\,e\,\frac{4ag^{2}_{\rm T}-bg^{2}_{\rm
V}}{6\,\pi^{2}}\,D^{\mu\nu}(x)\,A_{\mu}(x)\,D^{\dagger}_{\nu}(x)\nonumber\\
&+&i\,e\,\frac{g^{2}_{\rm
V}}{6\,\pi^{2}}\,(2\,b\,+\,3)\,D^{\dagger}_{\mu}(x)\,
D_{\nu}(x)\,F^{\mu\nu}(x)\nonumber\\
&+&i\,e\,(1\,+\,a)\,\frac{2\,g^{2}_{\rm
T}}{3\,\pi^{2}}\,\frac{1}{M^{2}_{\rm
D}}\,D^{\dagger}_{\mu\nu}(x)\,D^{\nu\lambda}(x)
\,{F_{\lambda}}^{\mu}(x)\,,
\end{eqnarray}
where $F_{\mu\nu}(x) = \partial_{\mu} A_{\nu}(x) - \partial_{\nu}
A_{\mu}(x)$ is the electromagnetic field strength, $a$ and $b$ are
arbitrary parameters related to the ambiguities of the one--nucleon
loop diagrams with respect to a shift of virtual nucleon momentum. We
consider them as free parameters of the approach [2].

In order to fix parameters it is convenient to write down the total
effective Lagrangian of the physical deuteron coupled to an external
electromagnetic field
\begin{eqnarray}\label{label4.11}
{\cal L}^{\rm D}_{\rm ELM}(x)_{\rm eff}&=&
-\frac{1}{2}\,D^{\dagger}_{\mu\nu}(x)D^{\mu\nu}(x) + M^2_{\rm
D}\,D^{\dagger}_{\mu}(x)D^{\mu}(x)\nonumber\\
&+&i\,e\,\frac{4ag^{2}_{\rm T}-bg^{2}_{\rm
V}}{6\,\pi^{2}}\frac{1}{6\,\pi^{2}}\,
D^{\dagger}_{\mu\nu}(x)\,A^{\mu}(x)\,D^{\nu}(x)\nonumber\\
&-&i\,e\,\frac{4ag^{2}_{\rm T}-bg^{2}_{\rm
V}}{6\,\pi^{2}}\,D^{\mu\nu}(x)\,A_{\mu}(x)\,D^{\dagger}_{\nu}(x)\nonumber\\
&+&i\,e\,\frac{g^{2}_{\rm
V}}{6 \pi^{2}}\,(2\,b\,+\,3)\,D^{\dagger}_{\mu}(x)\,
D_{\nu}(x)\,F^{\mu\nu}(x)\nonumber\\
&+&i\,e\,(1\,+\,a)\,\frac{2 g^{2}_{\rm
T}}{3 \pi^{2}}\,\frac{1}{M^{2}_{\rm
D}}\,D^{\dagger}_{\mu\nu}(x)\,D^{\nu\lambda}(x)
\,{F_{\lambda}}^{\mu}(x).
\end{eqnarray}
Two terms having the structure
$D^{\mu\nu}(x)A_{\mu}(x)D^{\dagger}_{\nu}(x)$ and
$D^{\dagger}_{\mu\nu}(x)A^{\mu}(x)D^{\nu}(x)$ should describe the
interaction of the deuteron with an external electromagnetic field
included by a minimal way, whilst the last two terms are responsible
for the non--trivial contributions to the magnetic dipole and electric
quadrupole moments of the deuteron.  In terms of the parameters of the
effective interactions Eq.(\ref{label4.11}) the magnetic dipole moment
$\mu_{\rm D}$, measured in nuclear magnetons, and the electric
quadrupole moment $Q_{\rm D}$, measure in ${\rm fm}^2$, of the
deuteron are given by
\begin{eqnarray}\label{label4.12}
\mu_{\rm D} &=&\frac{g^{2}_{\rm T}}{3\pi^{2}} +
(1\,+\,b)\,\frac{g^{2}_{\rm V}}{4\pi^{2}},\nonumber\\ 
Q_{\rm D} &=& \Bigg[(2\,+\,2\,a)\,\frac{g^{2}_{\rm T}}{3\pi^{2}} -
(3\,+\,2\,b)\,\frac{g^{2}_{\rm V}}{6\pi^{2}}\Bigg]\,\frac{1}{M^2_{\rm
D}}
\end{eqnarray}
at the constraint 
\begin{eqnarray}\label{label4.13}
b\,\frac{g^{2}_{\rm V}}{6\pi^{2}}\,-\,2\,a\,\frac{g^{2}_{\rm
 T}}{3\pi^{2}} = 1
\end{eqnarray}
reducing the first two terms in effective Lagrangian
Eq.(\ref{label4.10}) to the standard minimal form which can be
obtained from the effective Lagrangian of the free deuteron field by
the shift $\partial_{\mu}D_{\nu}(x) \to (\partial_{\mu} +
\,i\,e\,A_{\mu}(x)) D_{\nu}(x)$.

By retaining the former relation between the electric quadrupole
moment and the coupling constant $g_{\rm V}$, $Q_{\rm D} = 2g^2_{\rm
V}/\pi^2 M^2_{\rm D}$ [2], one can show that the experimental values
of the magnetic dipole moment of the deuteron $\mu_{\rm D} = 0.857$
and the electric quadrupole moment $Q_{\rm D} = 0.286\,{\rm fm}^2$ can
be fitted by the following values of the parameters $a$ and $b$: $a =
- \,0.442$ and $b = - \,4.418$ at $g_{\rm V} = 11.319$ [2]. This gives
the relation between the coupling constants $g_{\rm V}$ and $g_{\rm
T}$
\begin{eqnarray}\label{label4.14}
g_{\rm T} = -\,1.662\,g_{\rm V}.
\end{eqnarray}
Substituting this relation into Eq.(\ref{label4.8}) we can describe
the experimental value of the binding energy of the deuteron
$\varepsilon_{\rm D} = 2.225\,{\rm MeV}$ at the cut--off $\Lambda_{\rm
D} = 115.729\,{\rm MeV}$. The spatial region of virtual nucleon field
fluctuations forming the physical deuteron related to this value of
the cut--off $1/\Lambda_{\rm D} \sim \rho_{\rm D} = 1.705\,{\rm fm}$
agrees good with the effective range of the deuteron $\rho_{\rm D} =
(1.759\pm 0.005)\,{\rm fm}$ [6].

The effective Lagrangian of the deuteron field coupled to an external
electromagnetic field is given by
\begin{eqnarray}\label{label4.15}
\hspace{-0.5in}&&{\cal L}^{\rm D}_{\rm ELM}(x)_{\rm eff} =
-\frac{1}{2}\,[(\partial_{\mu} -
\,i\,e\,A_{\mu}(x))D^{\dagger}_{\nu}(x) - (\partial_{\nu} -
\,i\,e\,A_{\nu}(x))D^{\dagger}_{\mu}(x)]\nonumber\\
\hspace{-0.5in}&&\times\,[(\partial^{\mu} + \,i\,e\,A^{\mu}(x))D^{\nu}(x) -
(\partial^{\nu} + \,i\,e\,A^{\nu}(x))D^{\mu}(x)] + M^2_{\rm
D}\,D^{\dagger}_{\mu}(x)D^{\mu}(x)\nonumber\\ 
\hspace{-0.5in}&&-\,i\,e\,\bar{\mu}_{\rm D}\,D^{\dagger}_{\mu}(x)\,
D_{\nu}(x)\,F^{\mu\nu}(x)+\,i\,e\,\bar{Q}_{\rm
D}\,D^{\dagger}_{\mu\nu}(x)\,D^{\nu\lambda}(x)
\,{F_{\lambda}}^{\mu}(x),
\end{eqnarray}
where $\bar{\mu}_{\rm D} = 14.682\,\mu_{\rm D}$ and $\bar{Q}_{\rm D} =
0.514\,Q_{\rm D}$. The term of order $O(e^2)$ can be also derived in
the RFMD by using shift ambiguities of one--nucleon loop
diagrams. This term is required by the electromagnetic gauge
invariance of the effective Lagrangian of the deuteron field, but it
does not affect on the electromagnetic parameters of the deuteron
which are of order $O(e)$.

\section*{Conclusion}
\setcounter{equation}{0}

\hspace{0.2in} Unlike the first formulation of the RFMD given in
Refs.~[1,2], where we have stated that the RFMD {\it is far from being
induced by the dynamics of QCD and seems like an old--fashion
approach} [2], in this paper we have shown that the RFMD is in
complete agreement with QCD and can be formulated from the first
principles of QCD. The RFMD describes low--energy nuclear forces in
the nuclear phase of QCD in terms of one--nucleon loop
exchanges. One--nucleon loop exchanges provide a minimal way of the
transfer of nucleon flavours from an initial to a final nuclear state
and allow to take into account contributions of nucleon--loop
anomalies. These anomalies are related to high--energy fluctuations of
virtual nucleon fields, i.e. the $N\bar{N}$ fluctuations, and fully
determined by one--nucleon loop diagrams [33--35]. The dominance of
contributions of one--nucleon loop anomalies to effective Lagrangians
describing low--energy interactions of the deuteron coupled to itself
and other particles is justified in the large $N_C$ expansion in QCD
at $N_C \to \infty$. It is well--known that anomalies of quark--loop
diagrams play an important role for the correct description of strong
low--energy interactions of low--lying hadrons.  We argue an important
role of nucleon--loop anomalies for the correct description of
low--energy nuclear forces in the nuclear physics.

It is interesting that nucleon--loop anomalies can be interpreted as
non--trivial contributions of the non--perturbative quantum vacuum --
the nucleon Dirac sea [17].  In nuclear physics the influence of the
nucleon Dirac sea for low--energy properties of finite nuclei has been
analysed within quantum field theoretic approaches in the one--nucleon
loop approximation [16].  Unfortunately, in these approaches
contributions of one--nucleon loop anomalies have been taken into
account. The RFMD allows to fill this blank.

For the derivation of the RFMD from the first principles of QCD we
distinguish three non--perturbative phases of QCD: 1) the
low--energy quark--gluon phase (low--energy QCD), 2) the hadronic
phase and 3) the nuclear phase. Skipping over the intermediate
low--energy quark--gluon phase by means of the integration over high--
and low--energy quark and gluon fluctuations one arrives at the
hadronic phase of QCD containing only local hadron fields with quantum
numbers of mesons and baryons coupled at energies below the SB$\chi$S
scale $\Lambda_{\chi} \simeq 1\,{\rm GeV}$. The couplings of
low--lying mesons with masses less than the SB$\chi$S scale to
low--lying octet and decuplet of baryons can be described by Effective
Chiral Lagrangians with chiral $U(3)\times U(3)$ symmetry.

Integrating in the hadronic phase of QCD over heavy hadron degrees of
freedom with masses exceeding the SB$\chi$S scale one arrives at the
nuclear phase of QCD which characterizes itself by the appearance of
bound nucleon states -- nuclei. At low energies the result of
integration over heavy hadron degrees of freedom can be represented in
the form of phenomenological local many--nucleon interactions. Some of
these interactions are responsible for creation of many--nucleon
collective excitations which acquire the properties of observed nuclei
through nucleon--loop and low--lying meson exchanges.  This effective
field theory describes nuclei and processes of their low--energy
interactions by considering nuclei as elementary particles [28,29].

Following this scenario of the description of nuclei and their
low--energy interactions from the first principles of QCD the deuteron
should be produced in the nuclear phase of QCD by a phenomenological
local four--nucleon interaction as the Cooper np--pair with quantum
numbers of the deuteron. The properties of the physical deuteron,
i.e. the binding energy, the electric quadrupole moment and so, the
Cooper np--pair acquires through one--nucleon loop exchanges. As has
been shown the main part of the kinetic term of the effective
Lagrangian of the free physical deuteron field is induced by the
contribution of high--energy (short--distance) fluctuations of virtual
nucleon fields related to the anomaly of the one--nucleon loop diagram
with two vector vertices -- the VV--diagram.

In turn, the magnetic dipole $\mu_{\rm D}$ and electric quadrupole
moments of the physical deuteron $Q_{\rm D}$ are fully determined by
high--energy (short--distance) fluctuations of virtual nucleon fields
related to the anomalous contributions of the one--nucleon loop
diagrams with three vector vertices (the $\gamma VV$--diagram) [2].
Thus, high--energy (short--distance) fluctuations of nucleon fields
related to anomalies of one--nucleon loop diagrams play a dominant
role for the formation of the physical deuteron from the Cooper
np--pair.

As regards low--energy (long--distance) fluctuations of virtual
nucleon fields they give a significant contribution only to the
binding energy of the deuteron $\varepsilon_{\rm D}$. The strength of
low--energy (long--distance) fluctuations of virtual nucleon fields is
restricted by the cut--off $\Lambda_{\rm D} = 115.729\,{\rm MeV}$. The
spatial region of virtual nucleon field fluctuations forming the
physical deuteron related to this value of the cut--off
$1/\Lambda_{\rm D} \sim \rho_{\rm D} = 1.705\,{\rm fm}$ agrees good
with the effective range of the deuteron $\rho_{\rm D} = (1.759\pm
0.005)\,{\rm fm}$ [6].

It is well--known that in the potential model approach to the
description of the deuteron the electric quadrupole moment of the
deuteron $Q_{\rm D}$ is caused by nuclear tensor forces which play an
important role for the existence of the deuteron as a bound np--state.

The proportionality of the coupling constant of the phenomenological
local four--nucleon interaction Eq.(\ref{label3.3}), responsible for
creation of the Cooper np--pair with quantum numbers of the deuteron,
and the binding energy of the deuteron $\varepsilon_{\rm D}$
Eq.(\ref{label3.20}) to the electric quadrupole moment $Q_{\rm D}$
testifies an important role of nuclear tensor forces for the formation
of the deuteron in the RFMD.

To the evaluation of one--nucleon loop diagrams defining effective
Lagrangians describing processes of low--energy interactions of the
deuteron coupled to itself and other particles we apply expansions in
powers of the momenta of interacting particles and keep only leading
terms of the expansions [1--5]. This approximation can be justified in
the large $N_C$ expansion. Indeed, in QCD with the $SU(N_C)$ gauge
group at $N_C \to \infty$ the nucleon mass is proportional to the
number of quark colours [19]: $M_{\rm N} \sim N_C$.  Since for the
derivation of effective Lagrangians describing the deuteron and
amplitudes of low--energy nuclear processes all external momenta of
interacting particles should be kept off--mass shell, the masses of
virtual nucleon fields are larger compared with the external
momenta. An expansion of one--nucleon loop diagrams in powers of
$1/M_{\rm N}$ giving an external momentum expansion corresponds to the
expansion in powers of $1/N_C$. In this case the leading order in the
large $N_C$ expansion gives the leading order contributions in the
expansion in powers of external momenta of interacting particles.  We
should emphasize that anomalous contributions of one--nucleon loop
diagrams are determined by the least powers in external momentum
expansions. Thereby, the dominance of contributions of nucleon--loop
anomalies to effective Lagrangians describing low--energy nuclear
processes in the RFMD is fully supported by the large $N_C$
expansion. The accuracy of this approximation is rather high. Indeed,
the real parameter of the expansion of one--nucleon loop diagrams is
$1/M^2_{\rm N} \sim 1/N^2_C$ but not $1/M_{\rm N} \sim
1/N_C$. Thereby, next--to--leading corrections should be of order
$O(1/N^2_C)$.

The inclusion of the interaction of the deuteron field with the tensor
nucleon current Eq.(\ref{label4.1}) has given a possibility of the
self-consistent description of the electromagnetic properties of the
deuteron, the magnetic dipole moment $\mu_{\rm D}$ and the electric
quadrupole moment $Q_{\rm D}$, in terms of effective interactions of
the Corben--Schwinger and Aronson kinds induced by one--nucleon loop
diagrams [2]. By fitting the experimental values of the magnetic
dipole moment $\mu_{\rm D} = 0.857\,\mu_{\rm N}$, where $\mu_{\rm N} =
e/2M_{\rm N}$ is a nuclear magneton, and the electric quadrupole
moment $Q_{\rm D} = 0.286\,{\rm fm}^2$ supplemented by the requirement
of the electromagnetic gauge invariance of the effective Lagrangian of
the deuteron field coupled to an external electromagnetic field [2] we
have got the relation between the coupling constants $g_{\rm V}$ and
$g_{\rm T}$: $g_{\rm T} = -\,1.662\,g_{\rm V}$. Due to this relation
the experimental value of the binding energy of the deuteron can be
described by the cut--off $\Lambda_{\rm D} = 115.729\,{\rm MeV}$. This
corresponds to the spatial region of virtual nucleon field
fluctuations forming the physical deuteron $1/\Lambda_{\rm D} \sim
\rho_{\rm D} = 1.705\,{\rm fm}$ agreeing good with the effective range
of the deuteron $\rho_{\rm D} = (1.759\pm 0.005)\,{\rm fm}$ [6].  As
has been stated in Ref.[2] the contributions of the tensor nucleon
current Eq.(\ref{label4.1}) to the amplitudes of low--energy nuclear
processes of astrophysical interest such as the neutron--proton
radiative capture n + p $\to$ D + $\gamma$ for thermal neutrons and
the solar proton burning p + p $\to$ D + e$^+$ + $\nu_{\rm e}$ can be
neglected. The same statement is valid for the disintegration of the
deuteron by anti--neutrinos and neutrinos induced by charged
$\bar{\nu}_{\rm e}$ + D $\to$ e$^+$ + n + n, $\nu_{\rm e}$ + D $\to$
e$^-$ + p + p and neutral $\bar{\nu}_{\rm e}(\nu_{\rm e})$ + D $\to$
$\bar{\nu}_{\rm e}(\nu_{\rm e})$ + n + p weak currents, and the pep
process, p + e$^-$ + p $\to$ D + $\nu_{\rm e}$.

The quantum field theoretic scenario to treating nuclei as
many--nucleon collective excitations induced by phenomenological local
many--nucleon interactions allows a plain extension of the RFMD by the
inclusion of light nuclei ${^3}{\rm He}$, ${^3}{\rm H}$ and ${^4}{\rm
He}$ as three-- and four--nucleon collective excitations. The binding
energies  and other low--energy parameters  of these excitations should
be defined through nucleon--loop and low--lying meson exchanges. The
extension of the RFMD by the inclusion of ${^3}{\rm He}$, ${^3}{\rm
H}$ and ${^4}{\rm He}$ would give a possibility to continue
investigations of the reactions of the p--p chain [36] started with
the reaction p + p $\to$ D + e$^+$ + $\nu_{\rm e}$ and to apply the
extended version of the RFMD to the description of the reactions p + D
$\to$ ${^3}{\rm He}$ + $\gamma$, p + ${^3}{\rm He}$ $\to$ ${^4}{\rm
He}$ + e$^+$ + $\nu_{\rm e}$ and so on.

We would like to emphasize that Chiral perturbation theory can be
naturally incorporated into this effective field theoretic approach to
physics of low--energy interactions of nuclei in terms of Effective
Chiral Lagrangians with chiral $U(3)\times U(3)$ symmetry describing
low--lying baryons and mesons interacting at low energies
[37]. Unfortunately, the discussion of the inclusion of Chiral
perturbation theory into the RFMD and the description of low--energy
nuclear processes in the RFMD goes beyond the scope of this
paper. Therefore, we relegate readers to Refs.[37,38], where these
problems and a comparison of the RFMD with other approaches have been
discussed in details.

\section*{Acknowledgement}

\hspace{0.2in} The authors (A.N. Ivanov and N.I. Troitskaya) are
grateful to Prof. Randjbar--Daemi, the Head of the High Energy Section
of the Abdus Salam International Centre for Theoretical Physics (ICTP)
in Trieste, for warm and kind hospitality extended to them
during the whole period of their stay at ICTP, when this work was
started.


\newpage

\begin{thebibliography}{99}
\bibitem{[1]} 
A. N. Ivanov, N. I. Troitskaya, M. Faber and
H. Oberhummer, Phys. Lett. B361 (1995) 74.
\bibitem{[2]}
A. N. Ivanov, N. I. Troitskaya, M. Faber and H. Oberhummer,
Nucl. Phys. A617 (1997) 414 and Nucl.Phys. A625 (1997) 896 (Erratum).
\bibitem{[3]}
A. N. Ivanov, N. I. Troitskaya, M. Faber and H. Oberhummer,
Z. Phys. A358 (1997) 81.
\bibitem{[4]} 
A. N. Ivanov, H. Oberhummer, N. I. Troitskaya and M. Faber, 
{\it Solar proton burning, photon and anti--neutrino
disintegration of the deuteron in the relativistic field theory model
of the deuteron}, nucl--th/9810065, October 1998.
\bibitem{[5]} 
A. N. Ivanov, H. Oberhummer, N. I. Troitskaya and M. Faber, 
{\it Solar neutrino processes  in the relativistic field theory model
of the deuteron}, nucl--th/9811012, November 1998.
\bibitem{[6]} 
M. Kamionkowski and J. N. Bahcall,
ApJ. 420 (1994) 884.
\bibitem{[7]} 
E. G. Adelberger {\it et al.},
Rev. Mod. Phys. 70 (1998) 1265.
\bibitem{[8]}
R. Schiavilla {\it et al.},
Phys. Rev. C58 (1998) 1263.
\bibitem{[9]} 
S. Weinberg, 
Phys. Lett. B251 (1990) 288;
Nucl. Phys. B363 (1991) 3; Phys. Lett. B295 (1992) 114.
\bibitem{[10]} 
D. R. Kaplan, M. J. Savage and M. B. Wise,
Nucl. Phys. B478 (1996) 629; Phys. Lett. B424 (1998) 390;
Nucl. Phys. B534 (1998) 329; 
D. Kaplan, Nucl. Phys. B494 (1997) 471.
\bibitem{[11]}
M. J. Savage, K. A. Scaldeferri and M. B. Wise,
{\it {\rm n + p $\to$ d + $\gamma$} in Effective Field Theory},
nucl--th/9811029, November 1998.
\bibitem{[12]}
T.--S. Park, K. Kubodera, D.--P. Min and M. Rho,
ApJ. 507 (1998) 443;
X. Kong and F. Ravndal,
{\it Proton--Proton Fusion in Leading order of Effective Field Theory},
nucl--th/9902064, March 1999;
{\it Effective--Range Corrections to the Proton--Proton Fusion Rate},
nucl--th/9904066, April 1999.
\bibitem{[13]} 
K. Kubodera, 
{\it Chiral Symmetry in Nuclei},
nucl--th/9903057, March 1999 and references therein.
\bibitem{[14]} 
M. Butler and J.--W. Chen, {\it Elastic and Inelastic
Neutrino--Deuteron Scattering in Effective Field Theory},
nucl--th/9905059, June 1999.
\bibitem{[15]} 
P. A. M. Dirac,
Proc. Roy. Soc. A126 (1930) 360;
J. R. Oppenheimer,
Phys. Rev. 30 (1930) 939; 
J. D. Bjorken and S. D. Drell, 
in {\it RELATIVISTIC
QUANTUM MECHANICS}, McGraw--Hill, Inc., New York, 1964, pp.64--66 and
reference therein.
\bibitem{[16]}
J. D. Walecka,
Ann. Phys. (NY) 83 (1974) 121;
C. J. Horowitz and  B. D. Scrot,
Nucl. Phys. A368 (1981) 503;
Phys. Lett. B140 (1984) 181;
 R. J. Perry,
Phys. Lett. B182 (1986) 269;
 T. D. Cohen,
Phys. Rev. C45 (1992) 833; 
J. C. Caillon and J. Labarsouque,
Phys. Lett. B311 (1993) 19;
 J. Caro, E. Ruiz Arriola and  L. L. Salcedo,
Phys. Lett. B383 (1996) 9;
  M Matsuzaki,
Phys. Rev. C58 (1998) 3407.
\bibitem{[17]} 
R. Jackiw, 
{\it Field theoretic investigations in
current algebra}, {\it Topological investigations of quantized gauge
theories}, in {\it CURRENT ALGEBRA AND ANOMALIES}, 
S. B. Treiman,
R. Jackiw, B. Zumino and E. Witten (eds), World Scientific, Singapore,
p.81 and p.211; 
N. S. Manton, 
Ann. of Phys. (NY) 159 (1985) 220;
N. Ogawa, 
Progr. Theor. Phys. 90 (1993) 717; 
R. A. Bertlemann, 
in {\it
ANOMALIES IN QUANTUM FIELD THEORY}, Oxford Science Publications,
Clarendon Press--Oxford, 1996, pp.227--233 and references therein.
\bibitem{[18]}
G. 't Hooft,
Nucl. Phys. B75 (1974) 461.
\bibitem{[19]} 
E. Witten, 
Nucl. Phys. B160 (1979) 57; {\it The large
$1/N$ expansion in atomic and particle physics}, HUTP--79/A078, 1979.
\bibitem{[20]} 
K. Kikkawa, Progr. Theor. Phys. 56 (1976) 947;
H. Kleinert, {\it Proc. of Int. Summer School of Subnuclear Physics},
Erice 1976, Ed. A. Zichichi, p.289.  
\bibitem{[21]}
A. Dhar and S. R. Wadia,
Phys. Rev. Lett. 52 (1984); 
A. Dhar, R. Shankar and S. R. Wadia,
Phys. Rev. D31 (1985) 3256;
D. Ebert and H. Reinhart,
Nucl. Phys. B271 (1986) 188;
M. Wakamatsu, 
Ann. of Phys. (N.Y.) 193 (1989) 287.
\bibitem{[22]} 
A. N. Ivanov, M. Nagy and N. I. Troitskaya,
Int. J. Mod. Phys. A7 (1992) 7305; 
A. N. Ivanov, Int. J. Mod. Phys. A8
(1993) 853; 
A. N. Ivanov, N. I. Troitskaya and M. Nagy,
Int. J. Mod. Phys. A8 (1993) 2027, 3425; Phys. Lett. B295 (1992) 308;
Phys. Lett. B308 (1993) 111;
A. N. Ivanov and N. I. Troitskaya,
Nuovo Cimento A108 (1995) 555.
\bibitem{[23]}
J. Bijnens, C. Bruno and E. de Rafael,
Nucl. Phys. B390 (1993) 501;
J. Bijnens, E. de Rafael and H. Zheng,
Z. Phys. C62 (1994) 437.
\bibitem{[24]}
A. N. Ivanov, N. I. Troitskaya, M. Faber, M. Schaler and M. Nagy,
Nuovo Cim. A107 (1994) 1667; Phys. Lett. B336 (1995) 555;
A. N. Ivanov, N. I. Troitskaya and M. Faber,
Nuovo Cim. A108 (1995) 613.
\bibitem{[25]}
A. N. Ivanov, M. Nagy and N. I. Troitskaya,
Phys. Rev. C59 (1999) 451;
Ya. A. Berdnikov, A. N. Ivanov, V. F. Kosmach and N. I. Troitskaya,
Phys. Rev. C60 (1999) 015201.
\bibitem{[26]}
S. Gasiorowicz and D. A. Geffen,
Rev. Mod. Phys. 41 (1969) 531 and references therein.
\bibitem{[27]}
J. Wess and B. Zumino,
Phys. Lett. B37 (1971) 95.
\bibitem{[28]}
B. Sakita and C. J. Goebal,
Phys. Rev. 127 (1962) 1787;
B. Sakita,
Phys. Rev. 127 (1962) 1800.
\bibitem{[29]} 
C. W. Kim and H. Primakoff, 
{\it Nuclei as elementary
particles in weak and electromagnetic processes} in {\it MESONS IN
NUCLEI}, Vol.1 (1979) pp.67--106, ed. M. Rho and D. Wilkinson,
Noth--Holland Publishing Company Amsterdam--New York--Oxford;
C. W. Kim and H. Primakoff,
Phys. Rev. B139 (1965) 1447; {\it ibid.} B140 (1965) 586.
\bibitem{[30]}
M. M. Nagels {\it et al.},
Nucl. Phys. B147 (1979) 253.
\bibitem{[31]}
R. L. Jaffe, 
Phys. Rev. D15 (1977) 267, 281;
R. L. Jaffe and F. E. Low,
Phys. Rev. D19 (1979) 2105.
\bibitem{[32]}
N. N. Achasov, S. A. Devyanin and G. N. Shestakov,
Sov. J. Nucl. Phys. 32 (1980) 566; Phys. Lett. B96 (1980) 168;
Phys. Lett. B108 (1982) 134; Z. Phys. C16 (1982) 55;
N. N. Achasov and G. N. Shestakov,
Z. Phys. C41 (1988) 309;
N. N. Achasov and V. V. Gubin,
Phys. Rev. D56 (1997) 4084.
\bibitem{[33]}
I. S. Gertsein and R. Jackiw,
Phys. Rev. 1881 (1969) 1955.
\bibitem{[34]}
S. L. Adler and W. A. Bardeen,
Phys. Rev. 182 (1969) 1517.
\bibitem{[35]}
R. W. Brown, C. C. Shih and B. L. Yang,
Phys. Rev. 186 (1969) 1491.
\bibitem{[36]} 
C. E. Rolfs and W. S. Rodney, 
in {\it CAULDRONS IN THE
COSMOS}, the University of Chicago Press, Chicago and London, 1988.
\bibitem{[37]} 
A. N. Ivanov, H. Oberhummer, N. I. Troitskaya and M. Faber, 
{\it Neutron--proton radiative capture, photo--magnetic and
anti--neutrino disintegration of the deuteron in the relativistic
field theory model of the deuteron}, nucl--th/9908080, August  1999.
\bibitem{[38]} 
A. N. Ivanov, H. Oberhummer, N. I. Troitskaya and M. Faber, 
{\it Solar neutrino burning, neutrino disintegration of the
deuteron and pep process in the relativistic field theory model of the
deuteron}, nucl--th/9910021, October 1999.
\end{thebibliography}
\end{document}